\documentclass[draftclsnofoot,12pt, onecolumn]{IEEEtran}
\usepackage{amsmath}
\usepackage{epsfig}
\usepackage{amssymb}
\usepackage{latexsym}
\usepackage{graphicx}
\usepackage{cite}

\begin{document}
\title{Closed-Form Bounds for the Rice $Ie$-Function}
\author{\IEEEauthorblockN{Paschalis C. Sofotasios \\}
\IEEEauthorblockA{School of Electronic and Electrical Engineering \\
University of Leeds, UK\\
e-mail: p.sofotasios@leeds.ac.uk\\}
\and
\IEEEauthorblockN{Steven Freear\\}
\IEEEauthorblockA{School of Electronic and  Electrical Engineering \\
University of Leeds, UK\\
e-mail: s.freear@leeds.ac.uk}}
\maketitle
\begin{abstract} 
This work is devoted in the derivation of novel upper and lower bounds for the Rice $Ie$-function. These bounds are expressed in closed-form and are shown to be quite tight. This is particularly evident by the fact that for a certain range of parameter values, the derived lower bound virtually behaves as a remarkably accurate approximation. As a result, the offered expressions can be considered  useful mathematical tools that can be efficiently employed in various analytical studies related to natural sciences and engineering. To this effect, they can be sufficiently applied in the area of digital communications over fading channels for the derivation of explicit representations for vital performance measures such as bit and symbol error probability, among others.
\end{abstract}
%

%
\section{Introduction}
\indent
Over the last decades, special functions have played a crucial role in studies related to natural sciences and engineering. In the wide field of wireless communications, their application in various analytical studies often allows the derivation of closed-form expressions for critical performance measures such as probability of error, channel capacity and higher order statistics. Furthermore, the majority of special functions are found to be built-in in widely used mathematical software packages such as $Maple$, $Mathematica$ and $Matlab$. As a result, the algebraic representation of any associated relationships as well as their computational realization, have been simplified  enormously. \\
\indent
Among other, a special function related to the area of telecommunications is the Rice $Ie$-function. This is denoted as $Ie(k,x)$ and was firstly proposed by S. O. Rice about six decades ago \cite{1}. In the field of digital communications it has been largely utilized in the study of zero crossing rates. In addition, it has extensively appeared in the analysis of angle modulation systems as well as in radar pulse detection and error rates in differentially encoded systems, \cite{2,3,4,5}.  \\
\indent
The definition of the Rice $Ie$-function is typically given in an integral form with the involved integral being finite and comprising an exponential function and a modified Bessel function of the first kind and order zero. Alternative algebraic representations include two series expressions which were reported in \cite{4}. These series are infinite and are expressed in terms of the modified Struve function and the modified Bessel function of the first kind, \cite{6,7}. Moreover, the author in \cite{5} reported useful closed-form identities that relate the $Ie$-function with the Marcum Q-function, \cite{8,9,10, 11, 12, 13, New_1, New_2, New_3} and the references therein. \\
\indent
However, to the best of the authors' knowledge the Rice $Ie$-function is not explicitly expressed in terms of other special and/or elementary functions. In addition, it is not included as a built-in function in the aforementioned popular software packages. \\
\indent
Motivated by these statements, this work is devoted in deriving upper and lower bounds for the Rice $Ie$-function. These bounds are expressed in closed-form and are shown to be particularly tight. This is evident by the fact that for a certain range of parametric values, the lower bound virtually behaves as a remarkably accurate approximation. Furthermore, the algebraic form of the derived bounds is relatively simple and therefore, convenient to handle both analytically and numerically. This feature renders them useful for meaningful utilization in analytical studies relating to the area of digital communications over fading environments \cite{Additional_3, Additional_6, Additional_7, Additional_8, Additional_9} and the references therein.

The remainder of this paper is organised as follows: Section II revisits the definition of the $I_{e}(k,x)$ function along with its existing alternative representations. Novel upper and lower bounds are derived in Section III while their performance is analysed in Section IV. Finally, closing remarks are given in Section V. 
\section{The Rice $Ie$-function}
\indent
The Rice $Ie$-function is defined as \cite{4,5},

\begin{equation} \label{eq:one} 
Ie(k,x) \triangleq \int_{0}^{x} e^{-t} I_{0}(kt) dt
\end{equation}
which is valid for $0 \leq k \leq 1$. The notation $I_{0}(.)$ stands for the modified Bessel function of the first kind and zero order, \cite{6,7}. An equivalent integral representation to \eqref{eq:one} was proposed in \cite{4}, namely,

\begin{equation} \label{eq:two} 
Ie(k,x) = \frac{1}{\sqrt{1 - k^{2}}} - \frac{1}{\pi} \int_{0}^{\pi} \frac{e^{-x(1 - cos\theta)}}{1 - k cos\theta} d\theta
\end{equation}
In the same context, alternative representations for the $Ie(k,x)$ function were reported in \cite{4}. These representations are expressed in infinite series form and are given by, 
 
\begin{equation} \label{eq:three} 
Ie(k,x) = \sqrt{\frac{x \pi}{2 \sqrt{1 - k^{2}}}} e^{-x} \sum_{n = 0}^{\infty} \frac{1}{n!} \frac{x^{n} k^{2n}}{2^{n} \sqrt{1 - k^{2}}} \left[ \frac{1}{\sqrt{1 - k^{2}}} L_{n + \frac{1}{2}} \left(x\sqrt{1 - k^{2}} \right) + L_{n - \frac{1}{2}} \left(x\sqrt{1 - k^{2}} \right) \right]
\end{equation}
and

\begin{equation} \label{eq:four}  
\begin{split}
Ie(k,x) =& x e^{-x} \frac{\sqrt{\pi}}{2} \sum_{n = 0}^{\infty} \left( \frac{x\left(1 - k^{2} \right)}{2k} \right)^{n+1} \frac{I_{n+1}(kx)}{\Gamma \left(n + \frac{5}{2}  \right)}\\
&+ x e^{-x} \left[I_{0}(kx) + \frac{\sqrt{\pi}}{2k} \sum_{n = 0}^{\infty} \left( \frac{x\left(1 - k^{2} \right)}{2k} \right)^{n} \frac{I_{n+1}(kx)}{\Gamma \left(n + \frac{3}{2} \right)} \right]
\end{split}
  \end{equation}
where the notations $L_{n}(.)$, $I_{n}(.)$ and $\Gamma(.)$ denote, the modified Struve function, the modified Bessel function of the first kind and the Euler's gamma function,  respectively \cite{6,7}. According to \cite{4}, the $Ie$-function can be computed by using the above two series alternatively. To this effect, equation \eqref{eq:three} converges relatively quickly for the case that the term $x\sqrt{1 - k^{2}}$ is large while the term $kx$ is small. On the contrary, equation \eqref{eq:four} converges quickly when the term $x\sqrt{1 - k^{2}}$ is small while the term $kx$ is large. \\
\indent
Nevertheless, it is noted that this method for computing the Rice $Ie$-function is quite inefficient. The reasons underlying this statement are the following: $i)$ two relationships, instead of one, are required for the computation; $ii)$ the above series are relatively unstable due to their infinite form; $iii)$ the $L_{n}(.)$ function is not a built-in function in popular mathematical software packages such as $Maple$, $Mathematica$ and $Matlab$. \\
\indent
An adequate way to resolve was reported in \cite{5}. There, the Rice $Ie$-function is related to the Marcum Q-function by the following two expressions

\begin{equation} \label{eq:five} 
Ie(k,x) = \frac{1}{\sqrt{1 - k^{2}}} \left[2Q(a , b) - e^{-x}I_{0}(kx) - 1 \right]
\end{equation} 
and

\begin{equation} \label{eq:six} 
Ie(k,x) = \frac{1}{\sqrt{1 - k^{2}}} \left[ Q(a, b) - Q(b, a) \right]
\end{equation}
where, 

\begin{equation}
a=\sqrt{x}\sqrt{1 + \sqrt{1 - k^{2}}},
\end{equation}
and

\begin{equation}
b=\sqrt{x}\sqrt{1 - \sqrt{1 - k^{2}}}
\end{equation}
The notation $Q(a,b)$ denotes the Marcum Q-function of the first order which is defined as, \cite{9, 10}

\begin{equation} \label{eq:seven} 
Q(a,b) = Q_{1}(a,b) \triangleq \int_{b}^{\infty} x e^{-\frac{a^{2} + x^{2}}{2}} I_{0}(ax) dx
\end{equation}
\section{Bounds to the Rice $Ie$-function}
\indent
It is recalled here that the authors in [11] established the monotonicity criteria for the $Q_{m}(a,b)$ function and subsequently derived tight upper and lower bounds to it. By following a similar procedure in this Section, novel upper and lower bounds for the Rice $Ie$-function are derived in closed-form. To this end, it is noted that a primary requirement is the derivation of an alternative representation of $Ie$-function. 
 
\subsection{An alternative representation of the Rice $Ie$-function}
 
\textit{\textbf{Lemma 1.} For $x>0$ and $0 \leq k \leq 1$, the following relationship holds,} 

\begin{equation} \label{eq:eight} 
 Ie(k,x) = 1 - e^{-x}I_{0}(kx) + k \int_{0}^{x} e^{-t} I_{1}(kt)dt
\end{equation}
\noindent
\textit{\textbf{Proof.}} By integrating once by parts \eqref{eq:one}, it follows  immediately that 
 
\begin{equation} \label{eq:nine} 
Ie(k,x) = \left[ \int_{0}^{x} e^{-t} dt \right] I_{0}(kt) - \int_{0}^{x} \left[\int_{0}^{x} e^{-t} dt \right]\left[\frac{d}{dt} I_{0}(kt) \right] dt
\end{equation}
According to the basic principles of integration, the first integral in \eqref{eq:nine} is evaluated straightforwardly as, 

\begin{equation} \label{eq:ten} 
\int_{0}^{x} e^{-t} dt = -e^{-x}
\end{equation}
Regarding the derivative of the modified Bessel function of the first kind and order $n$, it is recalled that according to \cite{6, 7}, it follows that 

\begin{equation} \label{eq:eleven} 
\frac{d}{dx} I_{n}(kx) =\frac{k}{2} \left[ I_{n-1}(kx) + I_{n+1}(kx) \right]
\end{equation}
which for $x=t$ and $n=0$ yields,

\begin{equation} \label{eq:twelve} 
\frac{d}{dt} I_{0}(kt) = kI_{1}(kt)
\end{equation}
Subsequently, by substituting \eqref{eq:ten} and \eqref{eq:twelve} in \eqref{eq:nine} and recalling that by definition $I_{0}(0) \triangleq 1$, equation \eqref{eq:eight} is finally deduced and thus, the proof is completed. $\blacksquare$
\begin{figure}[h]
\centerline{\psfig{figure=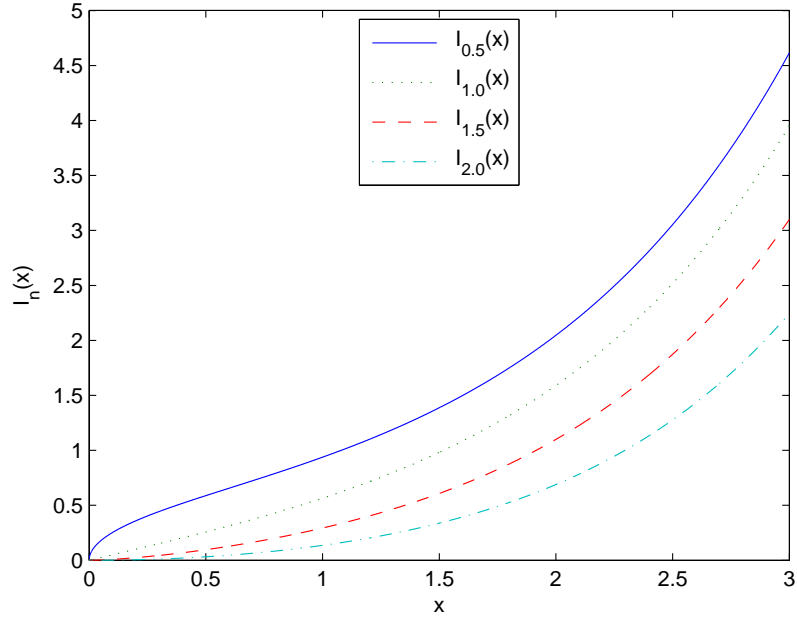, height=9cm, width=12cm}}
\caption{The monotonically decreasing behaviour of $I_{n}(x)$ w.r.t $n$.}
\end{figure}
\subsection{An upper bound for the Rice $Ie$-function} 
$ $\\
\noindent 
\textit{\textbf{Theorem 1.}} \textit{For $x>0$ and $0 \leq k \leq 1$, the following inequality holds,}

\begin{equation} \label{eq:thirteen} 
Ie(k,x) < 1 - e^{-x} I_{0}(kx) + \sqrt{\frac{k}{2}} \left[ \frac{erf(c\sqrt{x})}{c} - \frac{erf( d \sqrt{x})}{d} \right]
\end{equation}
where $c = \sqrt{1 - k}$ and $d = \sqrt{1 + k}$. 
\\
$ $
\\
\noindent
\textit{\textbf{Proof.}} It is recalled that the modified Bessel function of the first kind is strictly decreasing with respect to its order $n$. This is clearly illustrated in Figure $1$. To this effect, it is easily shown that the inequality $I_{n+a}(x) < I_{n}(x)$ holds for $a>0$. Based on this, the following inequality applies to \eqref{eq:eight}, 

\begin{equation} \label{eq:fourteen} 
Ie(k,x) < 1 - e^{-x}I_{0}(kx) + k \int_{0}^{x} e^{-t} I_{\frac{1}{2}}(kt)dt 
\end{equation}
Notably, for the special case that $n +0.5 \in \mathbb{N}$, the $I_{n}(x)$ function can be equivalently expressed in closed-form according to \cite[eq. (8.467)]{6}, namely,

\begin{equation} \label{eq:fifteen} 
I_{n + \frac{1}{2}}(x) \triangleq \sum_{k=0}^{n}\frac{(n+k)!\,\left[(-1)^{k}e^{x} + (-1)^{n+1}e^{-x}\right]}{\sqrt{\pi}k!(n-k)!(2x)^{k+\frac{1}{2}}}, \, n\in \mathbb{N}
\end{equation}
Therefore, for the special case where $n=0$, it follows that

\begin{equation} \label{eq:sixteen} 
I_{\frac{1}{2}}(kt) = \frac{e^{kt} - e^{-kt}}{\sqrt{2\pi kt}} 
\end{equation}
Evidently, the derivation of a closed-form upper bound to $Ie(k,x)$ is subject to evaluation in closed-form of the integral in \eqref{eq:fourteen}. To this end, with the aid of \eqref{eq:sixteen} it follows that

\begin{equation} \label{eq:seventeen} 
\int_{0}^{x} e^{-t} I_{\frac{1}{2}}(kt)dt = \int_{0}^{x} e^{-t} \left[\frac{e^{kt} - e^{-kt}}{\sqrt{2\pi kt}}\right] dt 
\end{equation}
which after long algebraic manipulations yields

\begin{equation} \label{eq:eighteen} 
\int_{0}^{x} e^{-t} I_{\frac{1}{2}}(kt)dt = \sqrt{\frac{k}{2}} \left[ \frac{erf(c\sqrt{x})}{c} - \frac{erf( d \sqrt{x})}{d} \right]
\end{equation}
where $c = \sqrt{1 - k}$ and $d = \sqrt{1 + k}$. Also,

\begin{equation}
erf(x) \triangleq \frac{2}{\sqrt{\pi}} \int_{0}^{x} e^{-t^{2}} dt
\end{equation}
\begin{figure}[h]
\centerline{\psfig{figure=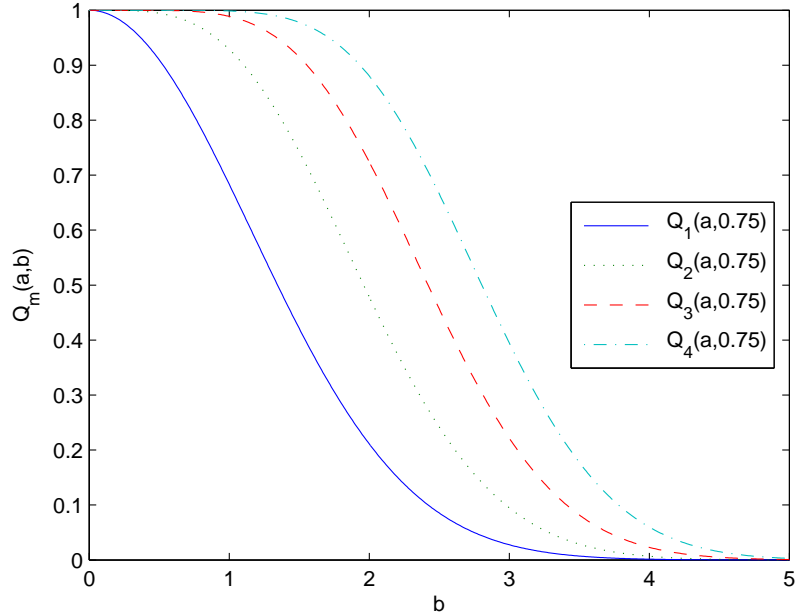, height=9cm, width=12cm}}
\caption{The monotonically increasing behaviour of $Q_{m}(a,b)$ w.r.t $m$.}
\end{figure}
denotes the error function \cite{6, 7}. Finally, by substituting \eqref{eq:eighteen} into \eqref{eq:fourteen}, equation \eqref{eq:thirteen} is deduced, which completes the proof. $\blacksquare$
\\
$ $
\\
\noindent
\textit{\textbf{Remark.}} The authors in \cite{11} derived closed-form bounds for the generalised Marcum Q-function, $Q_{m}(a,b)$. Therefore, at a first sight it appears that by making the necessary change of variables and substituting accordingly in \eqref{eq:five} and/or \eqref{eq:six}, an alternative expression for \eqref{eq:thirteen} can be deduced. Nevertheless, it is noted here that the corresponding resulted expression is clearly less compact and elegant compared to equation \eqref{eq:thirteen}.
\subsection{A lower bound for the Rice $Ie$-function} 
$ $ \\
\noindent 
\textit{\textbf{Theorem 2.}} \textit{For $x>0$ and $0 \leq k \leq 1$, the following inequality holds,}
$ $
\\
\begin{equation} \label{eq:nineteen} 
Ie(k,x)>\frac{2Q(b+a)+2Q(b-a) - e^{-x} I_{0}(kx) - 1}{\sqrt{1-k^{2}}}
\end{equation}
where
\begin{figure}[h]
\centerline{\psfig{figure=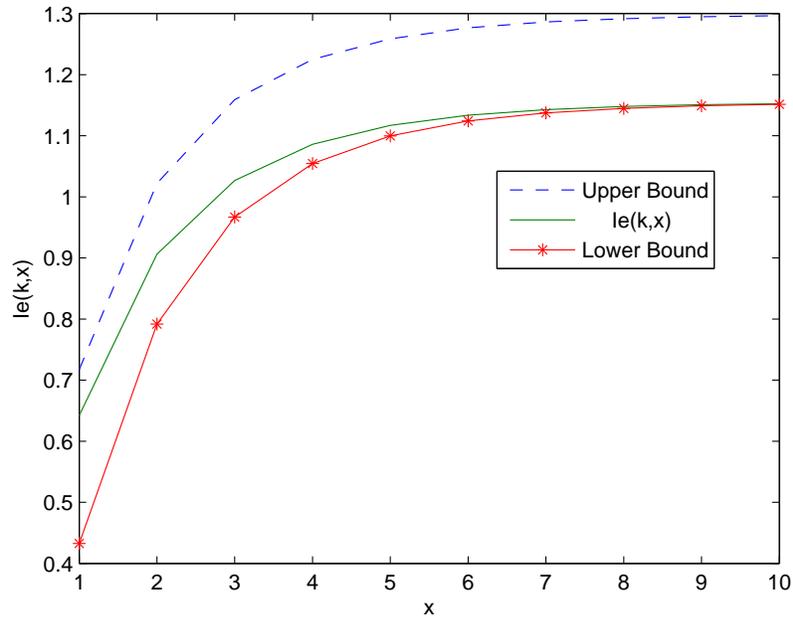, height=9cm, width=12cm}}
\caption{Behaviour of the bounds in eqs. $(13)$ and $(19)$ for $k=0.5$}
\end{figure}
\begin{equation}
Q(x) \triangleq \frac{1}{\sqrt{2 \pi}} \int_{x}^{\infty} e^{-\frac{t^{2}}{2}} dt
\end{equation}
denotes the one dimensional Gaussian Q-function \cite{6}.\\
$ $\\
\noindent
\textit{\textbf{Proof.}} According to the aforementioned monotonicity property of the $I_{n}(x)$ function, it immediately follows that that $I_{\frac{3}{2}}(x) < I_{1}(x)$. As a result, by making the necessary change of variables and substituting in equation \eqref{eq:eight}, the following closed-form inequality is deduced,
\begin{figure}[h]
\centerline{\psfig{figure=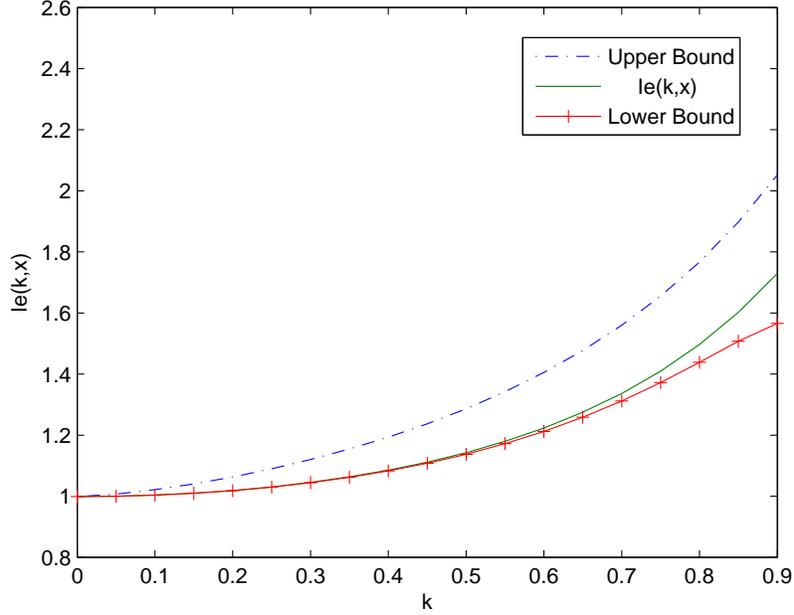, height=9cm, width=12cm}}
\caption{Behaviour of the bounds in eqs. $(13)$ and $(19)$ for $x=7$}
\end{figure}
\begin{equation} \label{eq:twenty} 
Ie(k,x) > 1 - e^{-x}I_{0}(kx) + k \int_{0}^{x} e^{-t} I_{\frac{3}{2}}(kt)dt 
\end{equation}

A similar inequality can be also deduced with the aid of equations \eqref{eq:five} and \eqref{eq:six}. To this effect, it is firstly recalled that the Marcum Q-function is strictly increasing with respect to the order $m$, which is clearly illustrated in Figure $2$. Based on this, the following inequality is obtained straightforwardly, 

\begin{equation} \label{eq:twentyone} 
Q_{1}(a,b)>Q_{\frac{1}{2}}(a,b)
\end{equation}
Subsequently, by substituting \eqref{eq:twentyone} into \eqref{eq:five}, it follows that

\begin{equation} \label{eq:twentytwo} 
Ie(k,x) > \frac{1}{\sqrt{1 - k^{2}}} \left[2Q_{\frac{1}{2}}(a, b) - e^{-x}I_{0}(kx) - 1 \right]
\end{equation}
It is recalled that based on the monotonicity criteria of the Marcum Q-function, the authors in [11] derived corresponding upper and lower bounds for $Q_{m}(a,b)$ along with a closed-form expression for the special case that $m$ is an odd multiple of $1/2$, i.e. $m+0.5 \in \mathbb{N}$. In this context, they thoroughly showed that 

\begin{equation} \label{eq:twentythree} 
Q_{\frac{1}{2}}(a,b) = Q(b+a) + Q(b-a)
\end{equation} 
To this effect, by substituting \eqref{eq:twentythree} into \eqref{eq:twentytwo} yields \eqref{eq:nineteen}, which completes the proof. $\blacksquare$

$ $
\\
\textit{\textbf{Remark.}} A lower bound for the $Ie(k,x)$ function could be theoretically derived by following the same methodology as in Theorem $1$. Nevertheless, this approach ultimately leads to an analytic expression that is divergent. 

\begin{figure}[h]
\centerline{\psfig{figure=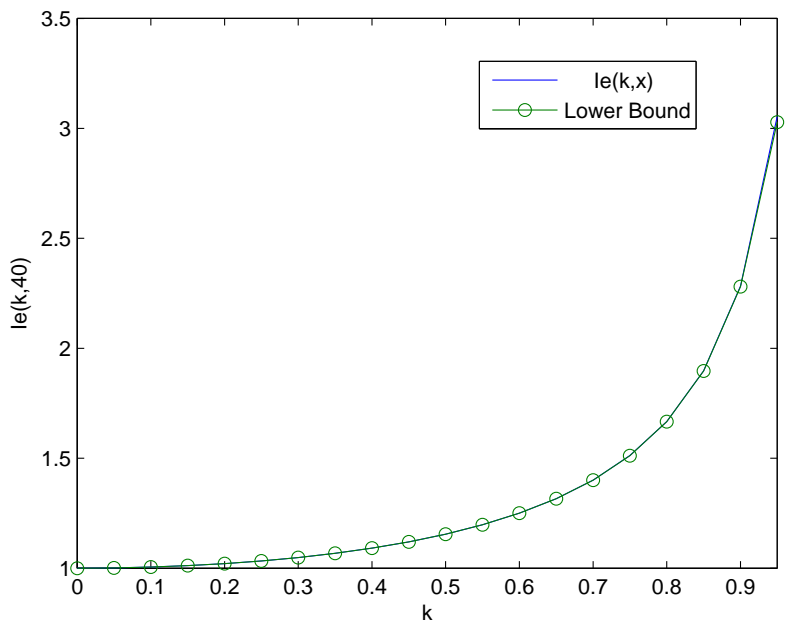, height=9cm, width=12cm}}
\caption{Behaviour of the lower bound in $(19)$ for $x=40$}
\end{figure}

\begin{figure}[h]
\centerline{\psfig{figure=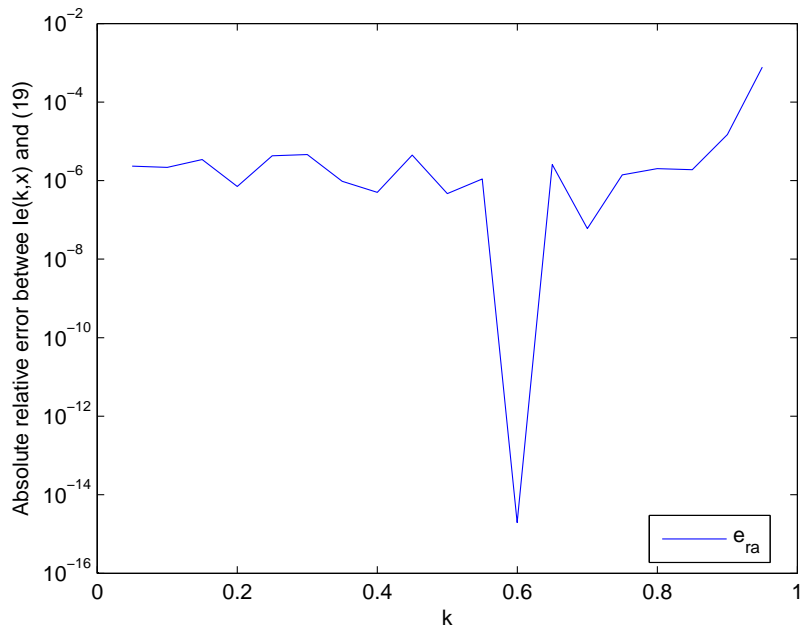, height=9cm, width=12cm}}
\caption{Absolute relative error, $\epsilon_{ar}$, between $Ie(k,x)$ and $(19)$ for $x=80$.}
\end{figure}
\section{Numerical Results}
\indent
This section investigates the tightness and overall behaviour of the derived closed-form bounds. To this end, Figure $3$ depicts the bounds in \eqref{eq:thirteen} and \eqref{eq:nineteen} with respect to $x$, for $k=0.5$ and in comparison with results obtained from numerical integrations. Evidently, it is shown that the upper bound is tighter than the lower bound for small values of $x$. However, it is noted that as $x$ increases, its tightness degrades while the lower bound becomes significantly tighter. As illustrated in Figure $4$, this behaviour also holds for relatively small values of $x$ as well as for different values of $k$. \\
\indent
Interestingly enough, for large values of $x$-typically $x>40$- the lower bound in \eqref{eq:nineteen} nearly becomes an exact representation for $Ie(k,x)$ as it appears to behave as a remarkably accurate approximation. This is illustrated in Figure $5$ where for the case that $x=80$, equation \eqref{eq:nineteen} provides an excellent match to the corresponding theoretical results. This is also justified by the low level of the involved absolute relative error which is given by $\epsilon_{ar} = \mid Ie(k,x) - eq. \eqref{eq:nineteen} \mid/ Ie(k,x)$, and is depicted in Figure $6$.
\section{Closing Remarks}
\indent
This work was devoted in the derivation of novel upper and lower bounds for the Rice $Ie$-function. These bounds are expressed in closed-form and are shown to be quite tight. This is particularly the case for the lower bound since for large values of the argument $x$, it becomes a remarkably accurate approximation to $Ie$-function over the whole range of values of $k$. As a result, it virtually behaves as an exact expression. Another advantage of the derived bounds is that their algebraic representation is relatively simple and therefore rather tractable both analytically and numerically. This is undoubtedly quite useful since the Rice $Ie$-function is not included as built-in function in widely mathematical software packages such as $Maple$, $Mathematica$ and $Matlab$. As a result, the offered expressions can be considered useful mathematical tools that can be meaningfully utilized in analytical studies related to the area of digital communications over fading channels. To this end, they can be indicatively applied efficiently in the derivation of explicit expressions for critical performance measures such as the bit and symbol error probability, among others.
\bibliographystyle{IEEEtran}
\end{document}